\documentclass[%
 reprint,
superscriptaddress,
 amsmath,amssymb,
 aps,%
]{revtex4-1}

\usepackage{graphicx}%
\usepackage{dcolumn}%
\usepackage{bm}%
\usepackage{float}
\usepackage{xcolor}
\usepackage{mhchem}
\usepackage{gensymb}
\usepackage{ulem}
\usepackage{comment}
\newcommand{\si}{Supplementary Materials}
\newcommand{\avg}[1]{\ensuremath{\left<#1\right>}}
\newcommand{\abs}[1]{\ensuremath{\lvert {#1}\rvert}}

\newcommand{\br}{\ensuremath{\textbf{r}}}
\newcommand{\bk}{\ensuremath{\textbf{k}}}

\newcommand{\bv}{\ensuremath{\textbf{v}}}
\newcommand{\bj}{\ensuremath{\textbf{j}}}

\begin{document}

\preprint{APS/123-QED}

\title{Computing the heat conductivity of fluids from density fluctuations}

\author{Bingqing Cheng}
\email{bc509@cam.ac.uk} 
\affiliation{
TCM Group, Cavendish Laboratory, University of Cambridge, J. J. Thomson Avenue, Cambridge CB3 0HE, United Kingdom
and \\
Trinity College, University of Cambridge, Cambridge CB2 1TQ, United Kingdom}%

\author{Daan Frenkel}
\affiliation{Department of Chemistry, University of Cambridge, Lensfield Road, Cambridge, CB2 1EW, United Kingdom}

\date{\today}%

\begin{abstract}
Equilibrium molecular dynamics simulations, in combination with the Green-Kubo (GK) method, have been extensively used to compute the thermal conductivity of liquids.
However, the GK method relies on an ambiguous definition of the microscopic heat flux, which depends on how one chooses to distribute energies over atoms. 
This ambiguity makes it problematic to employ the GK method for systems with non-pairwise interactions.
In this work, we show that  the hydrodynamic description of thermally driven density fluctuations  %
can be used to obtain the thermal conductivity of a bulk fluid unambiguously,  thereby bypassing the need to define the heat flux. 
We verify that,
for a model fluid with only pairwise interactions,
our method yields  estimates of thermal conductivity consistent with the GK approach.
We apply our approach to compute the thermal conductivity of a non-pairwise additive water model at supercritical conditions,
and then of a liquid hydrogen system described by a machine-learning interatomic potential, at 33 GPa and 2000 K.
\end{abstract}

\maketitle

\section{\label{sec:intro}Introduction}

The thermal conductivity $\lambda$ of a fluid measures how well it conducts heat.
Understanding the heat transport process is not only fundamentally important, but also has technological implications in material manufacturing, thermo-electric conversion~\cite{snyder2011complex}, energy saving, heat dissipation and many more. 

In insulators, heat transport is dominated by nuclear motion, as the electrons remain in the ground state and follow the nuclei adiabatically. 
Ever since the early days of Molecular Dynamics (MD), thermal conductivity computations for fluids have been of significant interest~\cite{Alder1970,levesque1973computer,evans1982homogeneous,vogelsang1987thermal}.
Some of these studies used non-equilibrium methods that impose a temperature gradient or a heat flux in the simulation box~\cite{evans1982homogeneous,muller1997simple,lukes2007thermal}.
The equilibrium methods typically exploit Green-Kubo (GK)
relations and rely on integrating equilibrium time correlation functions~\cite{green1954markoff,kubo1957statistical}.
For computing the thermal conductivity,
\begin{equation}
    \lambda = \dfrac{1}{V k_B T^2} 
    \int_0^\infty dt
    \avg{ \textbf{J}(0)\textbf{J}(t)},
    \label{eq:gk}
\end{equation}
where (for pairwise additive interactions) the heat flux $\textbf{J}(t) = \sum_i^N e_i \bv_i + \sum_{i<j} (\textbf{F}_{ij} \cdot \bv_i)\textbf{r}_{ij}$ is summed over all $N$ particles in the system, $e_i$ and $\bv_i$ are the atomic energy and velocity of atom $i$, $\textbf{F}_{ij}$ is the interaction force between atom $i$ and $j$ and $\textbf{r}_{ij}$ is their relative displacement vector.
The expression for $\textbf{J}$
is not unique - in particular for systems with non-pairwise additive interactions. 
To be more precise, 
although the integral in Eqn.~\eqref{eq:gk} at the infinite time limit does not depend on the specific choice of $\textbf{J}$ ~\cite{marcolongo2016microscopic},
the statistical accuracy of the calculation does depend on it~\cite{marcolongo2020gauge}.
For MD simulations involving many-body interactions~\cite{fan2015force}, such as density functional theory or machine learning potentials (MLPs)~\cite{behler2007generalized,bartok2010gaussian,cheng2019ab}, the partitioning of atomic energy and defining effective pairwise forces is ambiguous~\cite{fan2015force,Boone2019}.
This has consequences for simulations. 
For instance, a recent study~\cite{Boone2019} points out that the virial-stress heat flux expression used in LAMMPS~\cite{plimpton1993fast} (a popular MD code) cannot be extended to many-body potentials: ignoring this issue introduces systematic error in thermal conductivity predictions.
Furthermore, the GK formula in Eqn.~\eqref{eq:gk} calls for integrating to infinite time,
but in practice one has to truncate in $t$ to ensure the numerical integration is not dominated by statistical noise~\cite{ercole2017accurate}.
If the GK integral has a slowly decaying contribution, such truncation may introduce a systematic bias.

Here we 
explore an alternative method for computing  the thermal conductivity of liquids, which relies exclusively on analyzing fluctuations in the particle density-- an unambiguously quantity. 
Our approach harks back to the classical hydrodynamic theory that describes the density fluctuations responsible for Rayleigh-Brillouin scattering~\cite{hansen2013theory}. 
We benchmark the method on a simple model system over a wide range of temperature and density conditions.
Finally, we demonstrate applications on (i) water described by a monoatomic model~\cite{molinero2008water},
and (ii) high pressure hydrogen using a MLP~\cite{cheng2019evidence}.

\section{Theoretical background~\label{sec:method}}
We briefly recap the relevant theory (for more details, the reader is referred to Ref.~\cite{hansen2013theory} and the \si{}).
The particle-density field in a fluid  is defined as
\begin{equation}
    \rho(\br,t) = \sum_{i=1}^{N}
\delta[\br - \br_i(t)],
\label{eq:rho}
\end{equation}
where $\br_i(t)$ is the position of atom $i$ at time $t$.
For a system in a periodic cell with size $\{l_x,l_y,l_z\}$,
one can decompose the density field $\rho(\br,t)$ in discrete Fourier components, i.e.
\begin{multline}
    \widetilde{\rho}(\bk,t) = 
\dfrac{1}{V}\int_V d\br \rho(\br,t) e^{-i \bk \cdot \br}
=\dfrac{1}{V} \sum_{i=1}^{N} e^{-i \bk \cdot \br_i(t)}
,
\label{eq:rhokt}
\end{multline}
where $\bk = \{2\pi n_x/l_x,2\pi n_y/l_y,2\pi n_z/l_z\}$ is on the reciprocal lattice of the simulation cell.

In the hydrodynamic regime (wavelengths much larger than typical atomic dimensions), 
taking $\bk$ along the z-axis with $k=\abs{\bk}$,
one can write coupled hydrodynamic equations for the 
Fourier components of the density field $\widetilde{\rho}(\bk,t)$,
momentum density field $\widetilde{\bj}(\bk,t)$ and of the local temperature field $\widetilde{T}(\bk,t)$.
The full equations are given in Ref.~\cite{hansen2013theory} and in the \si.
Intuitively, the three quantities are coupled because density fluctuations may be driven by either temperature or pressure variations. %

The hydrodynamic equations can be solved approximately to the second order in $k$~\cite{hansen2013theory}:
transverse currents ($\widetilde{j}_x(\bk,t)$ and   $\widetilde{j}_y(\bk,t)$) have auto-correlation functions with an  exponential decay time determined by the fluid shear viscosity.
In the longitudinal direction, the approximate solution to $\widetilde{\rho}(\bk,t)$ has
two poles at
$s_0 = -iD_T k^2$
and
$s_{\pm} = \pm c_s k -i \Gamma k^2$.
Here 
$D_T=\lambda/c_P$ is thermal diffusivity,
$c_s$ is the adiabatic speed of sound,
the sound attenuation constant $\Gamma=(\gamma-1)D_T/2 + b/2$ where $b$ is the kinematic longitudinal viscosity,
and $\gamma = c_P/c_V$ is the ratio between volume-specific isobaric $c_P$ and isothermal $c_V$ heat capacities.
The auto-correlation function of the $\widetilde{\rho}(\bk,t)$ is
\begin{multline}
    C(\bk,t) = 
    \int_0^t dt 
    \avg{\widetilde{\rho}(\bk,0) \widetilde{\rho}(\bk,t)}\\
    =\rho_k^2 \left[
\dfrac{\gamma-1}{\gamma}\exp(-D_T k^2t)
+\dfrac{1}{\gamma}\exp(-\Gamma k^2 t) \cos (c_s k t)
\right],
\label{eq:cl}
\end{multline}
and its power spectrum is
\begin{multline}
    S(\bk,\omega) = \dfrac{S_k}{2\pi}\left[
\dfrac{\gamma-1}{\gamma}\dfrac{2 D_T k^2}{\omega^2+(D_T k^2)^2} \right.\\
\left. +\dfrac{1}{\gamma}\left(
\dfrac{\Gamma k^2}{(\omega+c_s k)^2+(\Gamma k^2)^2}
+\dfrac{\Gamma k^2}{(\omega-c_s k)^2+(\Gamma k^2)^2}
\right)\right].
\label{eq:sl}
\end{multline}
The above expressions are valid for small enough wavevector $k$  such that the sound velocity is determined by the {\em adiabatic} compressibility of the fluid (i.e. $k D_T \ll c_s$). 
The power spectrum (Eqn.~\eqref{eq:sl}) has three peaks:
the Rayleigh peak centered at the origin related to the diffusion of heat, and two Brillouin peaks at $\omega=\pm c_s k$,  due to propagating sound modes~\cite{hansen2013theory,forster2018hydrodynamic}.
The ratio between the area under the Rayleigh peak and the Brillouin peaks is $\gamma -1 $ (the Landau-Placzek ratio).
The width of the central peak is equal to $D_T k^2$ and hence proportional to $\lambda$. 
Systems with a larger $\gamma -1$ have a more prominent central peak, and are thus particularly suited for the computation of $\lambda$ with this method.

Eqn.~\eqref{eq:cl} and \eqref{eq:sl} are our key equations for computing $\lambda$:
We first Fourier expand the density field $\rho(\br,t)$ using Eqn.~\eqref{eq:rhokt}, from the MD trajectories generated at the constant volume and temperature (NVT) condition.
In practice, we fit the auto-correlation function or the power spectrum of $\widetilde{\rho}(\bk,t)$ to Eqn.~\eqref{eq:cl} or \eqref{eq:sl}, to estimate both $\lambda = c_P D_T$ and the kinematic longitudinal viscosity $b$.
We refer to this approach as the WAVE method.
The other quantities in Eqn.~\eqref{eq:cl} and \eqref{eq:sl} ($\gamma$ and $c_s$) can be computed separately and relatively easily (see \si), although one could choose to obtain $c_s$ also from the fit.

\section{Benchmark on a pairwise potential}

To validate the WAVE method,
we performed MD simulations using a Generalized Lennard-Jones (GLJ)  of Ref.~\cite{Wang2020} with a cut-off distance $r_{cut} = 2.0$ in reduced space units, and exponents $\nu=\mu=1$.
This GLJ is very similar to the LJ 12-6 potential, 
but avoids the ambiguity in the truncation at a finite cut-off radius.  
We performed NVT simulations for homogeneous fluid of 32,000 particles in a cubic simulation box, using the LAMMPS code~\cite{plimpton1993fast}.  
The global stochastic velocity rescaling thermostat~\cite{bussi2007canonical} was employed, as it scales the total kinetic energy of the system and is thus suitable for computing transport properties.
The time step was 0.0025 reduced time units, and the simulation length was 2,000,000 steps.
From the MD trajectories, we computed the time series of $\widetilde{\rho}(\bk,t)$ and fitted its auto-correlation function to Eqn.~\eqref{eq:sl} to estimate  $\lambda$ and $b$.
The power spectra of $\widetilde{\rho}(\bk,t)$ were calculated using a Fast Fourier Transform (see \si),
although cepstral analysis~\cite{ercole2017accurate} may be  statistically more efficient.
Because the GLJ~\cite{Wang2020} is a two-body potential, 
one can apply the GK method by defining $e_i$ using the equal splitting of all two-body energies and taking the readily available pairwise $\textbf{F}_{ij}$.

\begin{figure}
    \centering
    \includegraphics[width=0.45\textwidth]{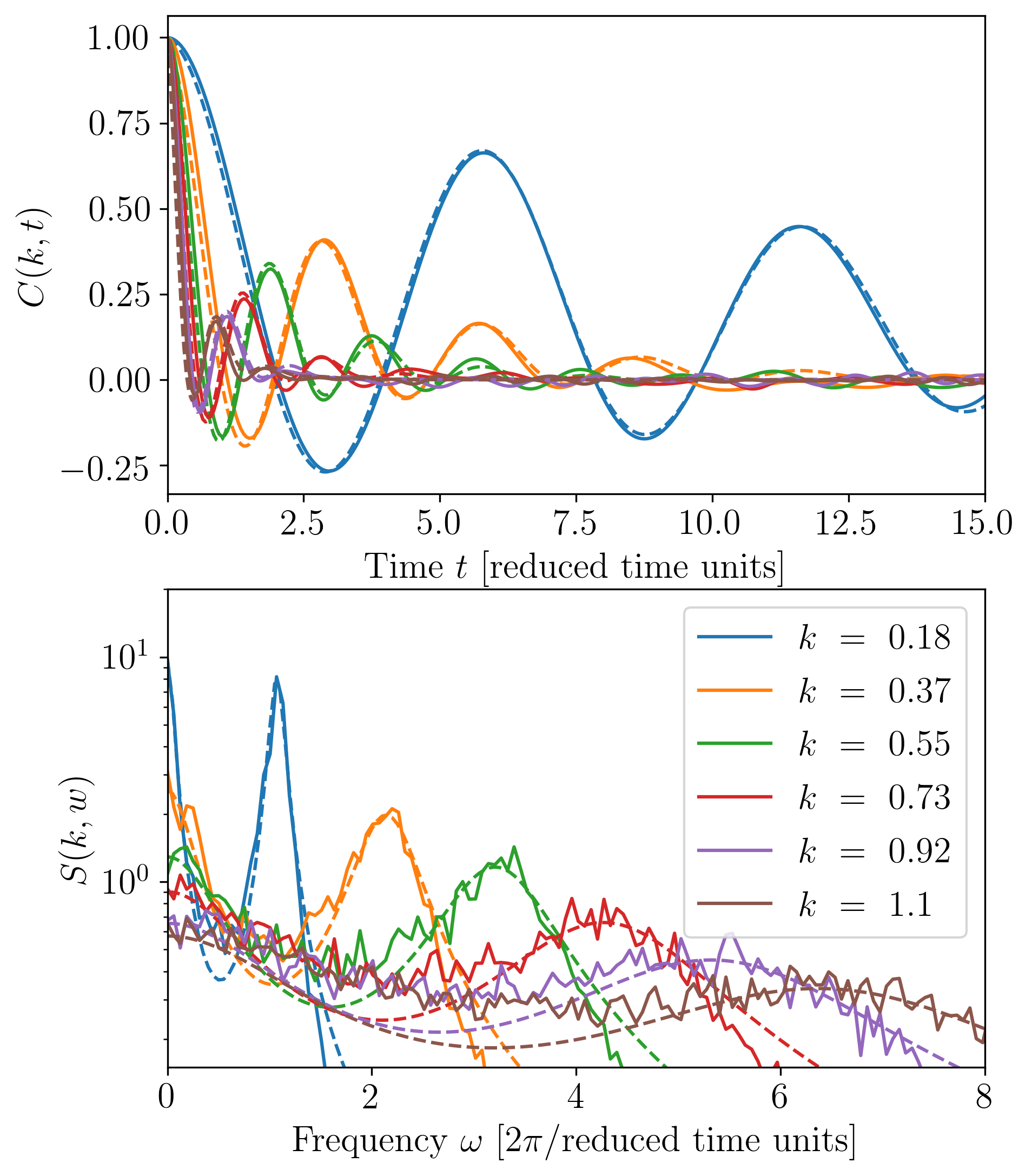}
    \caption{The auto-correlation functions (upper panel) and the power spectra (lower panel) of the real component of $\widetilde{\rho}(\bk,t)$ 
from an equilibrium molecular dynamics simulation using the GLJ potential~\cite{Wang2020} at $T=1.4$, $\rho=0.8$.
The solid curves show simulation results, and the dashed curves are the fits to Eqn.~\eqref{eq:cl} (upper panel) and Eqn.~\eqref{eq:sl} (lower panel).}
    \label{fig:fit}
\end{figure}
As an example, Fig.~\ref{fig:fit} displays the auto-correlation functions (upper panel) and the power spectra (lower panel) of $\widetilde{\rho}(\bk,t)$ obtained at $T=1.4$, $\rho=0.8$.
The solid curves indicate
simulation results at different $\bk$s %
along the $z$-axis of the simulation box.
The auto-correlations show oscillatory decay, and
the power spectra exhibit the characteristic Rayleigh and Brillouin peaks.
With $\lambda$ and $b$ as the two fitting parameter at each $\bk$,
Eqn.~\eqref{eq:cl} and ~\eqref{eq:sl} fit the simulation results well.

\begin{figure}
    \centering
    \includegraphics[width=0.40\textwidth]{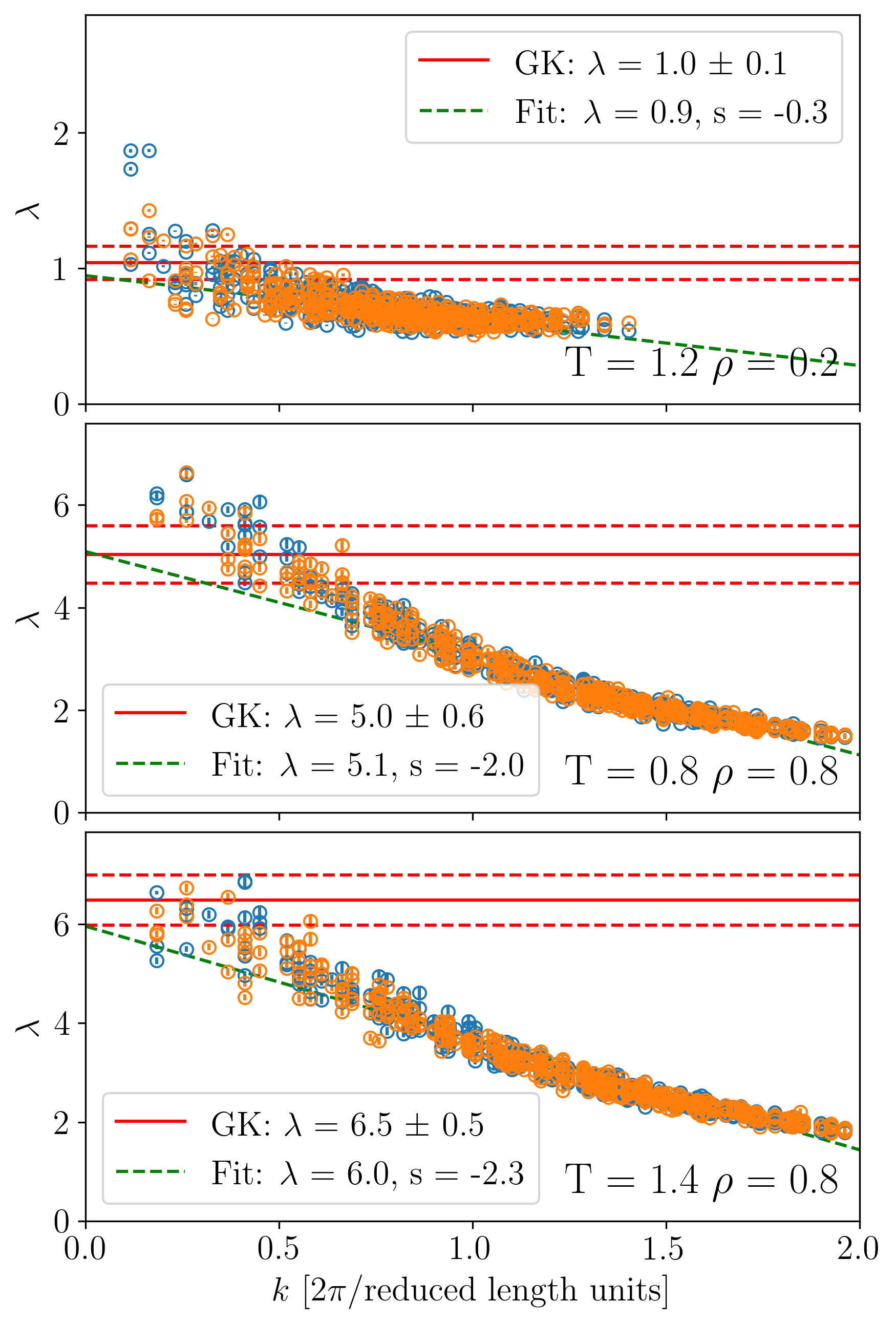}
    \caption{
    Thermal conductivities $\lambda$ 
from MD simulations of the GLJ liquids.
The solid red lines with error bands indicate the Green-Kubo values.
The blue (orange) dots show the fitted value of $\lambda(k)$ using the WAVE method from the real (imaginary) components of the density waves.
The dashed green lines indicate linear fits using $\lambda(k) = \lambda+s\times k$.} 
    \label{fig:all-results}
\end{figure}

We computed thermal conductivities over a wide range of conditions over a grid of temperature and density conditions ($0.2 \le \rho \le 0.8$, $0.8 \le T \le 2.0 $ in reduced units).
In Fig.~\ref{fig:all-results} we show
the three representative sets of results on low-density fluid (left panel), sub-critial liquid (middle panel), and supercritial liquid (right panel).
The remaining results for  20 other conditions are shown in the \si{}.
The values of $\lambda(k)$ at different $\bk$s were obtained,
with the largest $k$ corresponding to a wave length of a few atomic spacing.
$\lambda(k)$ is smooth in $k$, and we applied a linear extrapolation to $k=0$, to obtain the macroscopic thermal conductivity coefficients $\lambda$.
The statistical error in $\lambda$ estimated this way is small -- on the order of $10^{-2}$, but the systematic error in extrapolation may be larger.
We compared the predictions between the $\lambda$s from the GK and the WAVE method, and find good agreement
at all conditions considered.

The WAVE method also estimates the kinematic longitudinal viscosity $b$, related to the shear viscosity $\eta$ and the bulk viscosity $\zeta$ by  $b=(\frac{4}{3}\eta + \zeta)/\rho m$~\cite{hansen2013theory,forster2018hydrodynamic}.
One can compute $\eta$ from the transverse particle currents ($\widetilde{j}_x(\bk,t)$ and   $\widetilde{j}_y(\bk,t)$), using a method similar to WAVE~\cite{palmer1994transverse}.
In the \si, we show good agreement between such estimates of $\eta$ and the GK values.
However, we find a discrepancy between the GK and the WAVE methods for $b$, presumably coming from the bulk viscosity $\zeta$ contribution.
This discrepancy may either come from the GK side: the definition or the computation of the stress tensor, or a possible slowly decaying tail of its time correlation function, or the discrepancy may be due to the difficulty in extrapolating $\zeta(\bk)$ to the zero mode when using WAVE.

\section{Application to mW water}

\begin{figure}
    \centering
    \includegraphics[width=0.45\textwidth]{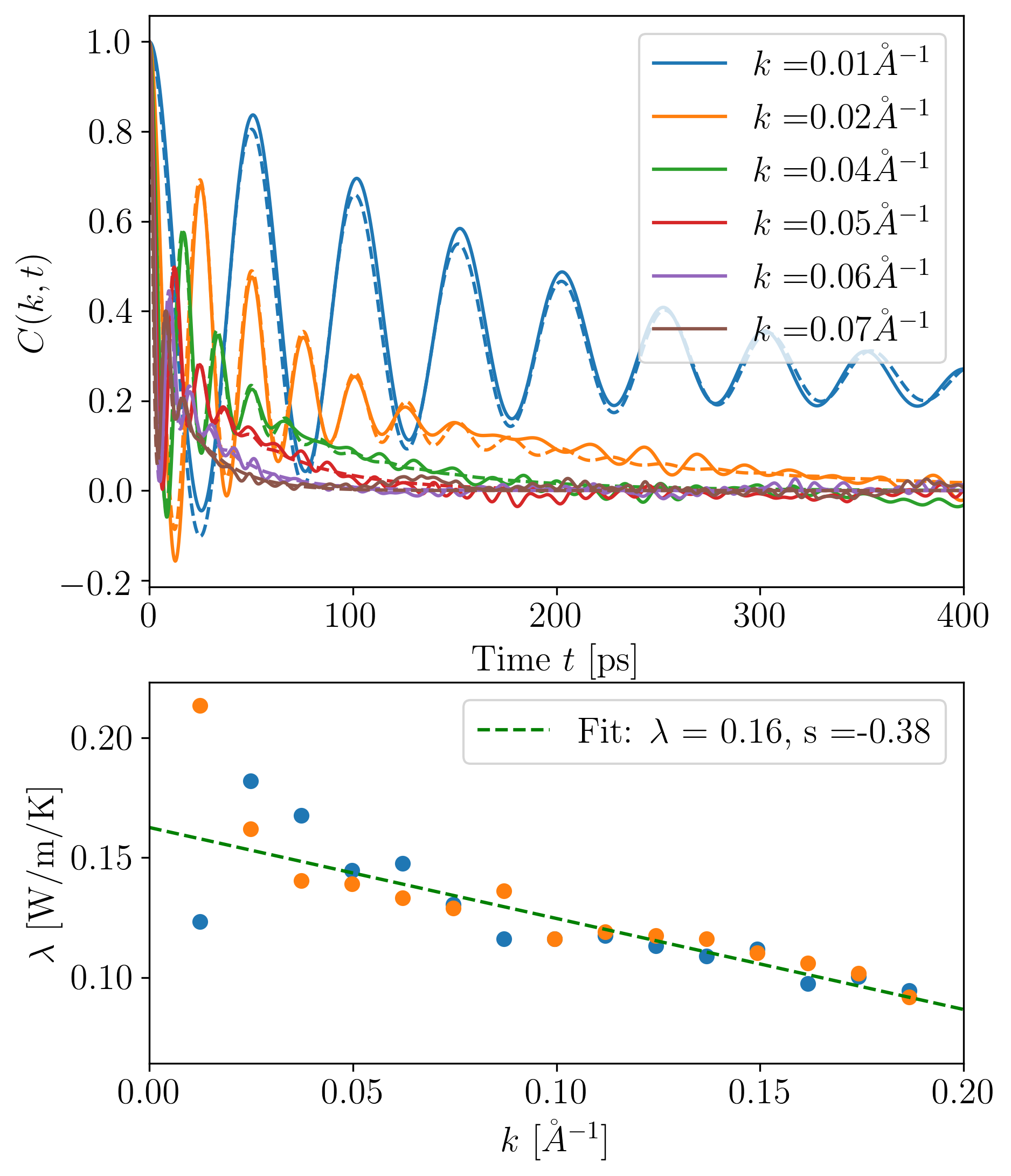}
    \caption{Upper: the auto-correlation functions of the real component of $\widetilde{\rho}(\bk,t)$
from an equilibrium MD simulation for mW water~\cite{molinero2008water} at 800~K and 355~atm.
The solid curves show simulation results,
and the dashed curves are the fits to Eqn.~\eqref{eq:cl}.
Lower: the blue and the orange dots show the fitted values of $\lambda$ from the real and the imaginary components, respectively.
The dashed green lines show a linear fit of $\lambda(k)$.}
    \label{fig:fit-mW}
\end{figure}

As an illustration of an application to a system described by a non-pairwise additive potential, we simulated a monoatomic water  model (mW)~\cite{molinero2008water}.
This model uses a Stillinger--Weber potential that combines two- and three-body interactions.
It is thus not straightforward to define the microscopic heat flux for this potential~\cite{fan2015force}.
We performed a NVT MD simulation at 800~K and 36~MPa (355~atm), which is in the supercritical region for water.
We used an elongated simulation box with 6,912 atoms, where the longest side has a length $l_z=50.5~\mathrm{nm}$.
The time step was 5~fs, and the simulation length was 10,000,000 steps.
We computed separately the volumetric specific heats $c_V=0.845(4)~\mathrm{J/K/cm^3}$ and $c_P=1.43(1)~\mathrm{J/K/cm^3}$, but left $c_s$ as a fitting parameter in Eqn.~\eqref{eq:cl}.
In Fig.~\ref{fig:fit-mW} 
we show the auto-correlation functions of the density waves (upper panel) for $\bk$s along the longest axis of the box, and the individual fitted values of $\lambda(k)$ (lower panel).
After a linear extrapolation to the zero $k$, we obtained
$\lambda=0.16~\mathrm{W/mK}$, $\Gamma=7\times10^{-7}~\mathrm{m}^2 / \mathrm{s}$ and $c_s=995~ \mathrm{m}/\mathrm{s}$.
This heat conductivity agrees fairly well with experimental data on supercritical water~\cite{mokry2009supercritical}. %
Knowledge of the thermal conductivity of water is relevant for modelling supercritical fluid-flow applications, e.g. in power engineering~\cite{mokry2011development}.

\section{Application to high pressure hydrogen}

\begin{figure}
    \centering
    \includegraphics[width=0.45\textwidth]{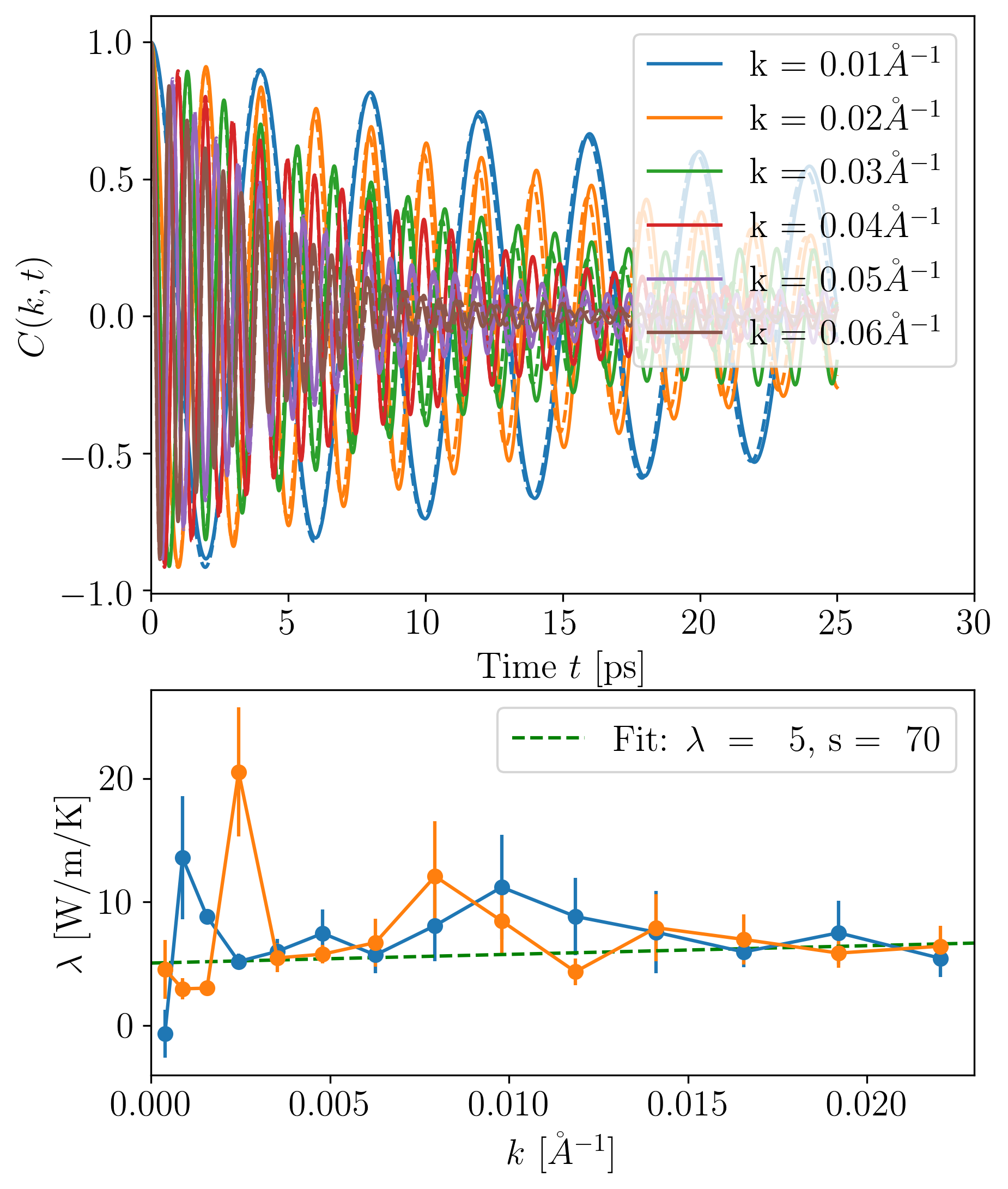}
    \caption{Upper: the auto-correlation functions of the real component of $\widetilde{\rho}(\bk,t)$
from an equilibrium MD simulation for dense liquid hydrogen~\cite{cheng2019evidence} at 2000~K and 33 GPa.
The solid curves show simulation results,
and the dashed curves are the fits to Eqn.~\eqref{eq:cl}.
Lower: the blue and the orange dots show the fitted values of $\lambda(k)$ from the real and the imaginary components, respectively.
The dashed green lines show a fit of $\lambda(k)$.}
    \label{fig:fit-highP-H}
\end{figure}

To showcase the application of the WAVE method to systems with truly many-body interactions,  we simulated high-pressure liquid hydrogen using a recently developed MLP~\cite{cheng2019evidence}.
This MLP uses Behler-Parrinello symmetry functions~\cite{behler2007generalized} to describe each atomic environment,
which are then used as the inputs for an artificial neural network with three hidden layers~\cite{Singraber2019}.
The GK expression in Eqn.~\eqref{eq:gk} based on $\textbf{F}_{ij}$
seems irreconcilable with the use of many-body potentials that cannot be
expressed a sum of pairwise additive terms.
The simulation was performed at 2000~K and 33~GPa (0.43 g/mL).
We used an elongated box ($l_z=63.5~\mathrm{nm}$) consisting 16,000 atoms.
The time step was 0.5~fs, and the simulation lasted 500~ps.
We computed separately $c_V=10.78(4)~\mathrm{J/K/cm^3}$, $c_P=10.96(5)~\mathrm{J/K/cm^3}$, and $c_s=1.6(1)\times 10^4~\mathrm{m/s}$.
Fig.~\ref{fig:fit-mW} 
shows the auto-correlation functions of the density waves (upper panel) for $\bk$s along the z-axis, 
and the individual fitted values of $\lambda(k)$ (lower panel).
From a linear extrapolation in $k$, we obtain
$\lambda=5~\mathrm{W/mK}$, $\Gamma=1.5\times10^{-6}~\mathrm{m}^2 / \mathrm{s}$.
This thermal conductivity is on the same order with the estimate from the extrapolation of values measured for a pressurized hydrogen gas at below 100 MPa~\cite{Moroe2011}.
It is, however, worth noting that nuclear quantum effects should be appreciable for hydrogen systems\cite{morales2013nuclear,cheng2018hydrogen}. 
Besides, the electronic contribution to the hydrogen thermal conductivity under these thermodynamic conditions has been estimated to be on the order of $10^{-4}~\mathrm{W/mK}$~\cite{holst2011electronic}. 
The estimates of $\lambda$ can have implications for 
 understanding the structural evolution of giant planets including Jupiter and Saturn~\cite{guillot_interiors_2005, chabrier2007heat},
 as
dense hydrogen is the dominant component of  the interior of these planets.
Knowledge of $\lambda$ is also relevant for  interpreting compression experiments in diamond anvil cells~\cite{houtput2019finite}.

\section{Conclusions}

In this paper, we propose an alternative method (WAVE) for computing  the thermal conductivity of liquids.
The method relies on analyzing particle density fluctuations in space and time.
We tested the method on a model potential, and demonstrated it correctness by comparing the estimates of $\lambda$ with the corresponding GK values.
While the GK expression depends on the definition of atomic energies and pairwise forces, the WAVE method only requires a time series of atomic positions from MD simulations.
This means WAVE is attractive for analyzing fluids with complex interactions, including the systems described by many-body force fields,
machine-learning potentials~\cite{behler2007generalized,bartok2010gaussian,cheng2019ab}, or first principle methods based on quantum mechanics.
To show the versatility of the method,
we considered first a non-pairwise additive water model,
and then a model for fluid hydrogen fluid at high pressures, described by a neural network machine learning potential.

\textbf{Data availability}
 Sample input files,  all the necessary PYTHON data analysis scripts and detailed procedures of computing $c_V$, $c_P$ and $c_s$  are included in the \si{}.

\begin{acknowledgments}
BC thanks Stefano Baroni, Aleks Reinhardt, Gareth Tribello, Chris Pickard, Michele Ceriotti, and Robert Jack for stimulating discussions and reading an early draft of the manuscript.
BC acknowledges funding from Swiss National Science Foundation (Project P2ELP2-184408).
BC acknowledges the resources provided by
CSCS under Project ID s957, and by the Cambridge Tier-2 system funded by EPSRC Tier-2 capital grant EP/P020259/1.
\end{acknowledgments}


\begin{thebibliography}{35}%
\makeatletter
\providecommand \@ifxundefined [1]{%
 \@ifx{#1\undefined}
}%
\providecommand \@ifnum [1]{%
 \ifnum #1\expandafter \@firstoftwo
 \else \expandafter \@secondoftwo
 \fi
}%
\providecommand \@ifx [1]{%
 \ifx #1\expandafter \@firstoftwo
 \else \expandafter \@secondoftwo
 \fi
}%
\providecommand \natexlab [1]{#1}%
\providecommand \enquote  [1]{``#1''}%
\providecommand \bibnamefont  [1]{#1}%
\providecommand \bibfnamefont [1]{#1}%
\providecommand \citenamefont [1]{#1}%
\providecommand \href@noop [0]{\@secondoftwo}%
\providecommand \href [0]{\begingroup \@sanitize@url \@href}%
\providecommand \@href[1]{\@@startlink{#1}\@@href}%
\providecommand \@@href[1]{\endgroup#1\@@endlink}%
\providecommand \@sanitize@url [0]{\catcode `\\12\catcode `\$12\catcode
  `\&12\catcode `\#12\catcode `\^12\catcode `\_12\catcode `\%12\relax}%
\providecommand \@@startlink[1]{}%
\providecommand \@@endlink[0]{}%
\providecommand \url  [0]{\begingroup\@sanitize@url \@url }%
\providecommand \@url [1]{\endgroup\@href {#1}{\urlprefix }}%
\providecommand \urlprefix  [0]{URL }%
\providecommand \Eprint [0]{\href }%
\providecommand \doibase [0]{http://dx.doi.org/}%
\providecommand \selectlanguage [0]{\@gobble}%
\providecommand \bibinfo  [0]{\@secondoftwo}%
\providecommand \bibfield  [0]{\@secondoftwo}%
\providecommand \translation [1]{[#1]}%
\providecommand \BibitemOpen [0]{}%
\providecommand \bibitemStop [0]{}%
\providecommand \bibitemNoStop [0]{.\EOS\space}%
\providecommand \EOS [0]{\spacefactor3000\relax}%
\providecommand \BibitemShut  [1]{\csname bibitem#1\endcsname}%
\let\auto@bib@innerbib\@empty
%</preamble>
\bibitem [{\citenamefont {Snyder}\ and\ \citenamefont
  {Toberer}(2011)}]{snyder2011complex}%
  \BibitemOpen
  \bibfield  {author} {\bibinfo {author} {\bibfnamefont {G.~J.}\ \bibnamefont
  {Snyder}}\ and\ \bibinfo {author} {\bibfnamefont {E.~S.}\ \bibnamefont
  {Toberer}},\ }in\ \href@noop {} {\emph {\bibinfo {booktitle} {Materials for
  sustainable energy.}}}\ (\bibinfo  {publisher} {World Scientific},\ \bibinfo
  {year} {2011})\ pp.\ \bibinfo {pages} {101--110}\BibitemShut {NoStop}%
\bibitem [{\citenamefont {Alder}\ \emph {et~al.}(1970)\citenamefont {Alder},
  \citenamefont {Gass},\ and\ \citenamefont {Wainwright}}]{Alder1970}%
  \BibitemOpen
  \bibfield  {author} {\bibinfo {author} {\bibfnamefont {B.~J.}\ \bibnamefont
  {Alder}}, \bibinfo {author} {\bibfnamefont {D.~M.}\ \bibnamefont {Gass}}, \
  and\ \bibinfo {author} {\bibfnamefont {T.~E.}\ \bibnamefont {Wainwright}},\
  }\href {\doibase 10.1063/1.1673845} {\bibfield  {journal} {\bibinfo
  {journal} {The Journal of Chemical Physics}\ }\textbf {\bibinfo {volume}
  {53}},\ \bibinfo {pages} {3813} (\bibinfo {year} {1970})}\BibitemShut
  {NoStop}%
\bibitem [{\citenamefont {Levesque}\ \emph {et~al.}(1973)\citenamefont
  {Levesque}, \citenamefont {Verlet},\ and\ \citenamefont
  {K{\"u}rkijarvi}}]{levesque1973computer}%
  \BibitemOpen
  \bibfield  {author} {\bibinfo {author} {\bibfnamefont {D.}~\bibnamefont
  {Levesque}}, \bibinfo {author} {\bibfnamefont {L.}~\bibnamefont {Verlet}}, \
  and\ \bibinfo {author} {\bibfnamefont {J.}~\bibnamefont {K{\"u}rkijarvi}},\
  }\href@noop {} {\bibfield  {journal} {\bibinfo  {journal} {Physical Review
  A}\ }\textbf {\bibinfo {volume} {7}},\ \bibinfo {pages} {1690} (\bibinfo
  {year} {1973})}\BibitemShut {NoStop}%
\bibitem [{\citenamefont {Evans}(1982)}]{evans1982homogeneous}%
  \BibitemOpen
  \bibfield  {author} {\bibinfo {author} {\bibfnamefont {D.~J.}\ \bibnamefont
  {Evans}},\ }\href@noop {} {\bibfield  {journal} {\bibinfo  {journal} {Physics
  Letters A}\ }\textbf {\bibinfo {volume} {91}},\ \bibinfo {pages} {457}
  (\bibinfo {year} {1982})}\BibitemShut {NoStop}%
\bibitem [{\citenamefont {Vogelsang}\ \emph {et~al.}(1987)\citenamefont
  {Vogelsang}, \citenamefont {Hoheisel},\ and\ \citenamefont
  {Ciccotti}}]{vogelsang1987thermal}%
  \BibitemOpen
  \bibfield  {author} {\bibinfo {author} {\bibfnamefont {R.}~\bibnamefont
  {Vogelsang}}, \bibinfo {author} {\bibfnamefont {C.}~\bibnamefont {Hoheisel}},
  \ and\ \bibinfo {author} {\bibfnamefont {G.}~\bibnamefont {Ciccotti}},\
  }\href@noop {} {\bibfield  {journal} {\bibinfo  {journal} {The Journal of
  chemical physics}\ }\textbf {\bibinfo {volume} {86}},\ \bibinfo {pages}
  {6371} (\bibinfo {year} {1987})}\BibitemShut {NoStop}%
\bibitem [{\citenamefont {M{\"u}ller-Plathe}(1997)}]{muller1997simple}%
  \BibitemOpen
  \bibfield  {author} {\bibinfo {author} {\bibfnamefont {F.}~\bibnamefont
  {M{\"u}ller-Plathe}},\ }\href@noop {} {\bibfield  {journal} {\bibinfo
  {journal} {The Journal of chemical physics}\ }\textbf {\bibinfo {volume}
  {106}},\ \bibinfo {pages} {6082} (\bibinfo {year} {1997})}\BibitemShut
  {NoStop}%
\bibitem [{\citenamefont {Lukes}\ and\ \citenamefont
  {Zhong}(2006)}]{lukes2007thermal}%
  \BibitemOpen
  \bibfield  {author} {\bibinfo {author} {\bibfnamefont {J.~R.}\ \bibnamefont
  {Lukes}}\ and\ \bibinfo {author} {\bibfnamefont {H.}~\bibnamefont {Zhong}},\
  }\href {\doibase 10.1115/1.2717242} {\bibfield  {journal} {\bibinfo
  {journal} {Journal of Heat Transfer}\ }\textbf {\bibinfo {volume} {129}},\
  \bibinfo {pages} {705} (\bibinfo {year} {2006})}\BibitemShut {NoStop}%
\bibitem [{\citenamefont {Green}(1954)}]{green1954markoff}%
  \BibitemOpen
  \bibfield  {author} {\bibinfo {author} {\bibfnamefont {M.~S.}\ \bibnamefont
  {Green}},\ }\href@noop {} {\bibfield  {journal} {\bibinfo  {journal} {The
  Journal of Chemical Physics}\ }\textbf {\bibinfo {volume} {22}},\ \bibinfo
  {pages} {398} (\bibinfo {year} {1954})}\BibitemShut {NoStop}%
\bibitem [{\citenamefont {Kubo}(1957)}]{kubo1957statistical}%
  \BibitemOpen
  \bibfield  {author} {\bibinfo {author} {\bibfnamefont {R.}~\bibnamefont
  {Kubo}},\ }\href@noop {} {\bibfield  {journal} {\bibinfo  {journal} {Journal
  of the Physical Society of Japan}\ }\textbf {\bibinfo {volume} {12}},\
  \bibinfo {pages} {570} (\bibinfo {year} {1957})}\BibitemShut {NoStop}%
\bibitem [{\citenamefont {Marcolongo}\ \emph {et~al.}(2016)\citenamefont
  {Marcolongo}, \citenamefont {Umari},\ and\ \citenamefont
  {Baroni}}]{marcolongo2016microscopic}%
  \BibitemOpen
  \bibfield  {author} {\bibinfo {author} {\bibfnamefont {A.}~\bibnamefont
  {Marcolongo}}, \bibinfo {author} {\bibfnamefont {P.}~\bibnamefont {Umari}}, \
  and\ \bibinfo {author} {\bibfnamefont {S.}~\bibnamefont {Baroni}},\
  }\href@noop {} {\bibfield  {journal} {\bibinfo  {journal} {Nature Physics}\
  }\textbf {\bibinfo {volume} {12}},\ \bibinfo {pages} {80} (\bibinfo {year}
  {2016})}\BibitemShut {NoStop}%
\bibitem [{\citenamefont {Marcolongo}\ \emph {et~al.}(2020)\citenamefont
  {Marcolongo}, \citenamefont {Ercole},\ and\ \citenamefont
  {Baroni}}]{marcolongo2020gauge}%
  \BibitemOpen
  \bibfield  {author} {\bibinfo {author} {\bibfnamefont {A.}~\bibnamefont
  {Marcolongo}}, \bibinfo {author} {\bibfnamefont {L.}~\bibnamefont {Ercole}},
  \ and\ \bibinfo {author} {\bibfnamefont {S.}~\bibnamefont {Baroni}},\
  }\href@noop {} {\bibfield  {journal} {\bibinfo  {journal} {Journal of
  Chemical Theory and Computation}\ } (\bibinfo {year} {2020})}\BibitemShut
  {NoStop}%
\bibitem [{\citenamefont {Fan}\ \emph {et~al.}(2015)\citenamefont {Fan},
  \citenamefont {Pereira}, \citenamefont {Wang}, \citenamefont {Zheng},
  \citenamefont {Donadio},\ and\ \citenamefont {Harju}}]{fan2015force}%
  \BibitemOpen
  \bibfield  {author} {\bibinfo {author} {\bibfnamefont {Z.}~\bibnamefont
  {Fan}}, \bibinfo {author} {\bibfnamefont {L.~F.~C.}\ \bibnamefont {Pereira}},
  \bibinfo {author} {\bibfnamefont {H.-Q.}\ \bibnamefont {Wang}}, \bibinfo
  {author} {\bibfnamefont {J.-C.}\ \bibnamefont {Zheng}}, \bibinfo {author}
  {\bibfnamefont {D.}~\bibnamefont {Donadio}}, \ and\ \bibinfo {author}
  {\bibfnamefont {A.}~\bibnamefont {Harju}},\ }\href@noop {} {\bibfield
  {journal} {\bibinfo  {journal} {Physical Review B}\ }\textbf {\bibinfo
  {volume} {92}},\ \bibinfo {pages} {094301} (\bibinfo {year}
  {2015})}\BibitemShut {NoStop}%
\bibitem [{\citenamefont {Behler}\ and\ \citenamefont
  {Parrinello}(2007)}]{behler2007generalized}%
  \BibitemOpen
  \bibfield  {author} {\bibinfo {author} {\bibfnamefont {J.}~\bibnamefont
  {Behler}}\ and\ \bibinfo {author} {\bibfnamefont {M.}~\bibnamefont
  {Parrinello}},\ }\href@noop {} {\bibfield  {journal} {\bibinfo  {journal}
  {Physical review letters}\ }\textbf {\bibinfo {volume} {98}},\ \bibinfo
  {pages} {146401} (\bibinfo {year} {2007})}\BibitemShut {NoStop}%
\bibitem [{\citenamefont {Bart{\'o}k}\ \emph {et~al.}(2010)\citenamefont
  {Bart{\'o}k}, \citenamefont {Payne}, \citenamefont {Kondor},\ and\
  \citenamefont {Cs{\'a}nyi}}]{bartok2010gaussian}%
  \BibitemOpen
  \bibfield  {author} {\bibinfo {author} {\bibfnamefont {A.~P.}\ \bibnamefont
  {Bart{\'o}k}}, \bibinfo {author} {\bibfnamefont {M.~C.}\ \bibnamefont
  {Payne}}, \bibinfo {author} {\bibfnamefont {R.}~\bibnamefont {Kondor}}, \
  and\ \bibinfo {author} {\bibfnamefont {G.}~\bibnamefont {Cs{\'a}nyi}},\
  }\href@noop {} {\bibfield  {journal} {\bibinfo  {journal} {Physical review
  letters}\ }\textbf {\bibinfo {volume} {104}},\ \bibinfo {pages} {136403}
  (\bibinfo {year} {2010})}\BibitemShut {NoStop}%
\bibitem [{\citenamefont {Cheng}\ \emph
  {et~al.}(2019{\natexlab{a}})\citenamefont {Cheng}, \citenamefont {Engel},
  \citenamefont {Behler}, \citenamefont {Dellago},\ and\ \citenamefont
  {Ceriotti}}]{cheng2019ab}%
  \BibitemOpen
  \bibfield  {author} {\bibinfo {author} {\bibfnamefont {B.}~\bibnamefont
  {Cheng}}, \bibinfo {author} {\bibfnamefont {E.~A.}\ \bibnamefont {Engel}},
  \bibinfo {author} {\bibfnamefont {J.}~\bibnamefont {Behler}}, \bibinfo
  {author} {\bibfnamefont {C.}~\bibnamefont {Dellago}}, \ and\ \bibinfo
  {author} {\bibfnamefont {M.}~\bibnamefont {Ceriotti}},\ }\href@noop {}
  {\bibfield  {journal} {\bibinfo  {journal} {Proceedings of the National
  Academy of Sciences}\ }\textbf {\bibinfo {volume} {116}},\ \bibinfo {pages}
  {1110} (\bibinfo {year} {2019}{\natexlab{a}})}\BibitemShut {NoStop}%
\bibitem [{\citenamefont {Boone}\ \emph {et~al.}(2019)\citenamefont {Boone},
  \citenamefont {Babaei},\ and\ \citenamefont {Wilmer}}]{Boone2019}%
  \BibitemOpen
  \bibfield  {author} {\bibinfo {author} {\bibfnamefont {P.}~\bibnamefont
  {Boone}}, \bibinfo {author} {\bibfnamefont {H.}~\bibnamefont {Babaei}}, \
  and\ \bibinfo {author} {\bibfnamefont {C.~E.}\ \bibnamefont {Wilmer}},\
  }\href {\doibase 10.1021/acs.jctc.9b00252} {\bibfield  {journal} {\bibinfo
  {journal} {Journal of Chemical Theory and Computation}\ }\textbf {\bibinfo
  {volume} {15}},\ \bibinfo {pages} {5579} (\bibinfo {year}
  {2019})}\BibitemShut {NoStop}%
\bibitem [{\citenamefont {Plimpton}(1993)}]{plimpton1993fast}%
  \BibitemOpen
  \bibfield  {author} {\bibinfo {author} {\bibfnamefont {S.}~\bibnamefont
  {Plimpton}},\ }\href@noop {} {\emph {\bibinfo {title} {Fast parallel
  algorithms for short-range molecular dynamics}}},\ \bibinfo {type} {Tech.
  Rep.}\ (\bibinfo  {institution} {Sandia National Labs., Albuquerque, NM
  (United States)},\ \bibinfo {year} {1993})\BibitemShut {NoStop}%
\bibitem [{\citenamefont {Ercole}\ \emph {et~al.}(2017)\citenamefont {Ercole},
  \citenamefont {Marcolongo},\ and\ \citenamefont
  {Baroni}}]{ercole2017accurate}%
  \BibitemOpen
  \bibfield  {author} {\bibinfo {author} {\bibfnamefont {L.}~\bibnamefont
  {Ercole}}, \bibinfo {author} {\bibfnamefont {A.}~\bibnamefont {Marcolongo}},
  \ and\ \bibinfo {author} {\bibfnamefont {S.}~\bibnamefont {Baroni}},\
  }\href@noop {} {\bibfield  {journal} {\bibinfo  {journal} {Scientific
  reports}\ }\textbf {\bibinfo {volume} {7}},\ \bibinfo {pages} {1} (\bibinfo
  {year} {2017})}\BibitemShut {NoStop}%
\bibitem [{\citenamefont {Hansen}\ and\ \citenamefont
  {McDonald}(2013)}]{hansen2013theory}%
  \BibitemOpen
  \bibfield  {author} {\bibinfo {author} {\bibfnamefont {J.-P.}\ \bibnamefont
  {Hansen}}\ and\ \bibinfo {author} {\bibfnamefont {I.~R.}\ \bibnamefont
  {McDonald}},\ }\href@noop {} {\emph {\bibinfo {title} {Theory of simple
  liquids, 3rd Edition}}}\ (\bibinfo  {publisher} {Elsevier},\ \bibinfo {year}
  {2013})\BibitemShut {NoStop}%
\bibitem [{\citenamefont {Molinero}\ and\ \citenamefont
  {Moore}(2008)}]{molinero2008water}%
  \BibitemOpen
  \bibfield  {author} {\bibinfo {author} {\bibfnamefont {V.}~\bibnamefont
  {Molinero}}\ and\ \bibinfo {author} {\bibfnamefont {E.~B.}\ \bibnamefont
  {Moore}},\ }\href@noop {} {\bibfield  {journal} {\bibinfo  {journal} {The
  Journal of Physical Chemistry B}\ }\textbf {\bibinfo {volume} {113}},\
  \bibinfo {pages} {4008} (\bibinfo {year} {2008})}\BibitemShut {NoStop}%
\bibitem [{\citenamefont {Cheng}\ \emph
  {et~al.}(2019{\natexlab{b}})\citenamefont {Cheng}, \citenamefont {Mazzola},\
  and\ \citenamefont {Ceriotti}}]{cheng2019evidence}%
  \BibitemOpen
  \bibfield  {author} {\bibinfo {author} {\bibfnamefont {B.}~\bibnamefont
  {Cheng}}, \bibinfo {author} {\bibfnamefont {G.}~\bibnamefont {Mazzola}}, \
  and\ \bibinfo {author} {\bibfnamefont {M.}~\bibnamefont {Ceriotti}},\
  }\href@noop {} {\bibfield  {journal} {\bibinfo  {journal} {arXiv preprint
  arXiv:1906.03341}\ } (\bibinfo {year} {2019}{\natexlab{b}})}\BibitemShut
  {NoStop}%
\bibitem [{\citenamefont {Forster}(2018)}]{forster2018hydrodynamic}%
  \BibitemOpen
  \bibfield  {author} {\bibinfo {author} {\bibfnamefont {D.}~\bibnamefont
  {Forster}},\ }\href@noop {} {\emph {\bibinfo {title} {Hydrodynamic
  fluctuations, broken symmetry, and correlation functions}}}\ (\bibinfo
  {publisher} {CRC Press},\ \bibinfo {year} {2018})\BibitemShut {NoStop}%
\bibitem [{\citenamefont {Wang}\ \emph {et~al.}(2020)\citenamefont {Wang},
  \citenamefont {Ram{\'{\i}}rez-Hinestrosa}, \citenamefont {Dobnikar},\ and\
  \citenamefont {Frenkel}}]{Wang2020}%
  \BibitemOpen
  \bibfield  {author} {\bibinfo {author} {\bibfnamefont {X.}~\bibnamefont
  {Wang}}, \bibinfo {author} {\bibfnamefont {S.}~\bibnamefont
  {Ram{\'{\i}}rez-Hinestrosa}}, \bibinfo {author} {\bibfnamefont
  {J.}~\bibnamefont {Dobnikar}}, \ and\ \bibinfo {author} {\bibfnamefont
  {D.}~\bibnamefont {Frenkel}},\ }\href {\doibase 10.1039/c9cp05445f}
  {\bibfield  {journal} {\bibinfo  {journal} {Physical Chemistry Chemical
  Physics}\ } (\bibinfo {year} {2020}),\ 10.1039/c9cp05445f}\BibitemShut
  {NoStop}%
\bibitem [{\citenamefont {Bussi}\ \emph {et~al.}(2007)\citenamefont {Bussi},
  \citenamefont {Donadio},\ and\ \citenamefont
  {Parrinello}}]{bussi2007canonical}%
  \BibitemOpen
  \bibfield  {author} {\bibinfo {author} {\bibfnamefont {G.}~\bibnamefont
  {Bussi}}, \bibinfo {author} {\bibfnamefont {D.}~\bibnamefont {Donadio}}, \
  and\ \bibinfo {author} {\bibfnamefont {M.}~\bibnamefont {Parrinello}},\
  }\href@noop {} {\bibfield  {journal} {\bibinfo  {journal} {The Journal of
  chemical physics}\ }\textbf {\bibinfo {volume} {126}},\ \bibinfo {pages}
  {014101} (\bibinfo {year} {2007})}\BibitemShut {NoStop}%
\bibitem [{\citenamefont {Palmer}(1994)}]{palmer1994transverse}%
  \BibitemOpen
  \bibfield  {author} {\bibinfo {author} {\bibfnamefont {B.~J.}\ \bibnamefont
  {Palmer}},\ }\href@noop {} {\bibfield  {journal} {\bibinfo  {journal}
  {Physical Review E}\ }\textbf {\bibinfo {volume} {49}},\ \bibinfo {pages}
  {359} (\bibinfo {year} {1994})}\BibitemShut {NoStop}%
\bibitem [{\citenamefont {Mokry}\ \emph {et~al.}(2009)\citenamefont {Mokry},
  \citenamefont {Gospodinov}, \citenamefont {Pioro},\ and\ \citenamefont
  {Kirillov}}]{mokry2009supercritical}%
  \BibitemOpen
  \bibfield  {author} {\bibinfo {author} {\bibfnamefont {S.}~\bibnamefont
  {Mokry}}, \bibinfo {author} {\bibfnamefont {Y.}~\bibnamefont {Gospodinov}},
  \bibinfo {author} {\bibfnamefont {I.}~\bibnamefont {Pioro}}, \ and\ \bibinfo
  {author} {\bibfnamefont {P.}~\bibnamefont {Kirillov}},\ }in\ \href@noop {}
  {\emph {\bibinfo {booktitle} {International Conference on Nuclear
  Engineering}}},\ Vol.\ \bibinfo {volume} {43543}\ (\bibinfo {year} {2009})\
  pp.\ \bibinfo {pages} {747--754}\BibitemShut {NoStop}%
\bibitem [{\citenamefont {Mokry}\ \emph {et~al.}(2011)\citenamefont {Mokry},
  \citenamefont {Pioro}, \citenamefont {Farah}, \citenamefont {King},
  \citenamefont {Gupta}, \citenamefont {Peiman},\ and\ \citenamefont
  {Kirillov}}]{mokry2011development}%
  \BibitemOpen
  \bibfield  {author} {\bibinfo {author} {\bibfnamefont {S.}~\bibnamefont
  {Mokry}}, \bibinfo {author} {\bibfnamefont {I.}~\bibnamefont {Pioro}},
  \bibinfo {author} {\bibfnamefont {A.}~\bibnamefont {Farah}}, \bibinfo
  {author} {\bibfnamefont {K.}~\bibnamefont {King}}, \bibinfo {author}
  {\bibfnamefont {S.}~\bibnamefont {Gupta}}, \bibinfo {author} {\bibfnamefont
  {W.}~\bibnamefont {Peiman}}, \ and\ \bibinfo {author} {\bibfnamefont
  {P.}~\bibnamefont {Kirillov}},\ }\href@noop {} {\bibfield  {journal}
  {\bibinfo  {journal} {Nuclear Engineering and Design}\ }\textbf {\bibinfo
  {volume} {241}},\ \bibinfo {pages} {1126} (\bibinfo {year}
  {2011})}\BibitemShut {NoStop}%
\bibitem [{\citenamefont {Singraber}\ \emph {et~al.}(2019)\citenamefont
  {Singraber}, \citenamefont {Behler},\ and\ \citenamefont
  {Dellago}}]{Singraber2019}%
  \BibitemOpen
  \bibfield  {author} {\bibinfo {author} {\bibfnamefont {A.}~\bibnamefont
  {Singraber}}, \bibinfo {author} {\bibfnamefont {J.}~\bibnamefont {Behler}}, \
  and\ \bibinfo {author} {\bibfnamefont {C.}~\bibnamefont {Dellago}},\ }\href
  {\doibase 10.1021/acs.jctc.8b00770} {\bibfield  {journal} {\bibinfo
  {journal} {Journal of Chemical Theory and Computation}\ }\textbf {\bibinfo
  {volume} {15}},\ \bibinfo {pages} {1827} (\bibinfo {year}
  {2019})}\BibitemShut {NoStop}%
\bibitem [{\citenamefont {Moroe}\ \emph {et~al.}(2011)\citenamefont {Moroe},
  \citenamefont {Woodfield}, \citenamefont {Kimura}, \citenamefont {Kohno},
  \citenamefont {Fukai}, \citenamefont {Fujii}, \citenamefont {Shinzato},\ and\
  \citenamefont {Takata}}]{Moroe2011}%
  \BibitemOpen
  \bibfield  {author} {\bibinfo {author} {\bibfnamefont {S.}~\bibnamefont
  {Moroe}}, \bibinfo {author} {\bibfnamefont {P.~L.}\ \bibnamefont
  {Woodfield}}, \bibinfo {author} {\bibfnamefont {K.}~\bibnamefont {Kimura}},
  \bibinfo {author} {\bibfnamefont {M.}~\bibnamefont {Kohno}}, \bibinfo
  {author} {\bibfnamefont {J.}~\bibnamefont {Fukai}}, \bibinfo {author}
  {\bibfnamefont {M.}~\bibnamefont {Fujii}}, \bibinfo {author} {\bibfnamefont
  {K.}~\bibnamefont {Shinzato}}, \ and\ \bibinfo {author} {\bibfnamefont
  {Y.}~\bibnamefont {Takata}},\ }\href {\doibase 10.1007/s10765-011-1052-5}
  {\bibfield  {journal} {\bibinfo  {journal} {International Journal of
  Thermophysics}\ }\textbf {\bibinfo {volume} {32}},\ \bibinfo {pages} {1887}
  (\bibinfo {year} {2011})}\BibitemShut {NoStop}%
\bibitem [{\citenamefont {Morales}\ \emph {et~al.}(2013)\citenamefont
  {Morales}, \citenamefont {McMahon}, \citenamefont {Pierleoni},\ and\
  \citenamefont {Ceperley}}]{morales2013nuclear}%
  \BibitemOpen
  \bibfield  {author} {\bibinfo {author} {\bibfnamefont {M.~A.}\ \bibnamefont
  {Morales}}, \bibinfo {author} {\bibfnamefont {J.~M.}\ \bibnamefont
  {McMahon}}, \bibinfo {author} {\bibfnamefont {C.}~\bibnamefont {Pierleoni}},
  \ and\ \bibinfo {author} {\bibfnamefont {D.~M.}\ \bibnamefont {Ceperley}},\
  }\href@noop {} {\bibfield  {journal} {\bibinfo  {journal} {Physical review
  letters}\ }\textbf {\bibinfo {volume} {110}},\ \bibinfo {pages} {065702}
  (\bibinfo {year} {2013})}\BibitemShut {NoStop}%
\bibitem [{\citenamefont {Cheng}\ \emph {et~al.}(2018)\citenamefont {Cheng},
  \citenamefont {Paxton},\ and\ \citenamefont {Ceriotti}}]{cheng2018hydrogen}%
  \BibitemOpen
  \bibfield  {author} {\bibinfo {author} {\bibfnamefont {B.}~\bibnamefont
  {Cheng}}, \bibinfo {author} {\bibfnamefont {A.~T.}\ \bibnamefont {Paxton}}, \
  and\ \bibinfo {author} {\bibfnamefont {M.}~\bibnamefont {Ceriotti}},\
  }\href@noop {} {\bibfield  {journal} {\bibinfo  {journal} {Physical review
  letters}\ }\textbf {\bibinfo {volume} {120}},\ \bibinfo {pages} {225901}
  (\bibinfo {year} {2018})}\BibitemShut {NoStop}%
\bibitem [{\citenamefont {Holst}\ \emph {et~al.}(2011)\citenamefont {Holst},
  \citenamefont {French},\ and\ \citenamefont {Redmer}}]{holst2011electronic}%
  \BibitemOpen
  \bibfield  {author} {\bibinfo {author} {\bibfnamefont {B.}~\bibnamefont
  {Holst}}, \bibinfo {author} {\bibfnamefont {M.}~\bibnamefont {French}}, \
  and\ \bibinfo {author} {\bibfnamefont {R.}~\bibnamefont {Redmer}},\
  }\href@noop {} {\bibfield  {journal} {\bibinfo  {journal} {Physical Review
  B}\ }\textbf {\bibinfo {volume} {83}},\ \bibinfo {pages} {235120} (\bibinfo
  {year} {2011})}\BibitemShut {NoStop}%
\bibitem [{\citenamefont {Guillot}(2005)}]{guillot_interiors_2005}%
  \BibitemOpen
  \bibfield  {author} {\bibinfo {author} {\bibfnamefont {T.}~\bibnamefont
  {Guillot}},\ }\href {\doibase 10.1146/annurev.earth.32.101802.120325}
  {\bibfield  {journal} {\bibinfo  {journal} {Annual Review of Earth and
  Planetary Sciences}\ }\textbf {\bibinfo {volume} {33}},\ \bibinfo {pages}
  {493} (\bibinfo {year} {2005})}\BibitemShut {NoStop}%
\bibitem [{\citenamefont {Chabrier}\ and\ \citenamefont
  {Baraffe}(2007)}]{chabrier2007heat}%
  \BibitemOpen
  \bibfield  {author} {\bibinfo {author} {\bibfnamefont {G.}~\bibnamefont
  {Chabrier}}\ and\ \bibinfo {author} {\bibfnamefont {I.}~\bibnamefont
  {Baraffe}},\ }\href@noop {} {\bibfield  {journal} {\bibinfo  {journal} {The
  Astrophysical Journal Letters}\ }\textbf {\bibinfo {volume} {661}},\ \bibinfo
  {pages} {L81} (\bibinfo {year} {2007})}\BibitemShut {NoStop}%
\bibitem [{\citenamefont {Houtput}\ \emph {et~al.}(2019)\citenamefont
  {Houtput}, \citenamefont {Tempere},\ and\ \citenamefont
  {Silvera}}]{houtput2019finite}%
  \BibitemOpen
  \bibfield  {author} {\bibinfo {author} {\bibfnamefont {M.}~\bibnamefont
  {Houtput}}, \bibinfo {author} {\bibfnamefont {J.}~\bibnamefont {Tempere}}, \
  and\ \bibinfo {author} {\bibfnamefont {I.~F.}\ \bibnamefont {Silvera}},\
  }\href@noop {} {\bibfield  {journal} {\bibinfo  {journal} {Physical Review
  B}\ }\textbf {\bibinfo {volume} {100}},\ \bibinfo {pages} {134106} (\bibinfo
  {year} {2019})}\BibitemShut {NoStop}%
\end{thebibliography}
\end{document}